\documentclass[aps,preprint]{revtex4}

\usepackage[dvips]{graphicx}

\newcommand{\ch}{{\cal H}}
\newcommand{\ce}{{\cal E}}

\begin{document}

\date{\today}
\title{Avoided level crossings in open quantum systems}

\author{Hichem Eleuch and Ingrid Rotter\footnote{rotter@pks.mpg.de}}
\address{Max Planck Institute for the Physics of Complex
Systems, D-01187 Dresden, Germany}

\begin{abstract}
At high level density, two states avoid usually crossing at the critical
value $a_{\rm cr}$ of the parameter $a$ by which the system is controlled.
The wavefunctions of the two
states are mixed in a finite parameter range around  $a_{\rm  cr}$.
This holds true for discrete states as well as for narrow resonance
states which are coupled via the environment of scattering wavefunctions.  
We study the influence of avoided level crossings onto 
four overlapping complex eigenvalues of a symmetric non-Hermitian operator.
The mixing of the two wavefunctions around $a_{\rm cr}$ is simulated,
in each case, by assuming a
Gaussian distribution around $a_{\rm cr}$. At high level density, the
Gaussian distributions related to avoided crossings of different
levels may overlap. Here, new effects arise, especially 
from the imaginary part of the coupling term via the environment.  
The results show, moreover, the influence of
symmetries onto the multi-level avoided crossing phenomenon.

\end{abstract}

\maketitle                   

\section{Introduction}
\label{intr}

Avoided level crossings play an important role in quantum mechanics. 
They are known for about 80 years \cite{landau}. In the case  
that the Hamilton operator of the system is Hermitian and
the real eigenvalues $E_i$  provide the energies of the 
discrete states of the system, two of its states  never will
cross. Instead they avoid crossing when controlled by a
parameter. This phenomenon is an observable effect: 
it consists in an exchange of the two states, including their
populations, in a many-particle system. 
The exchange takes place at the critical value $a_{\rm cr}$ of the 
parameter $a$. 
The corresponding crossing points of the energy trajectories can be 
found by analytical continuation into the continuum \cite{solov}.

Narrow resonance states show a similar behavior. They are described
well by the complex eigenvalues $E_i - i/2 \Gamma_i$ and
biorthogonal eigenfunctions $\Phi_i$ of a non-Hermitian Hamilton operator 
\cite{top}. The eigenvalues provide not only the energies $E_i$
but also the widths $\Gamma_i$
(inverse lifetimes) of the states. Mostly, the resonance states avoid
crossing similar to discrete states. However, in contrast to 
the eigenvalue trajectories of discrete states, the eigenvalue
trajectories of resonance states 
can cross. The crossing points are called usually {\it
  exceptional points} according to Kato \cite{kato} who studied first the
nontrivial mathematical properties of these singular points.
Also in the case of avoided  crossings of resonance states, the 
corresponding crossing points can be found by varying one additional
parameter. 

Recent studies have shown qualitatively that 
dynamical phase transitions in open quantum systems occur when the level
density is high and, correspondingly, many neighbored states avoid 
crossing  in a relatively small parameter range
\cite{top}. As a result, the system is (dynamically) stabilized:
the number of states is reduced (by one when the environment consists
of one continuum),  the individual spectroscopic
properties of the original states are lost, and the narrow
(trapped) resonance states of the system show collective features.
Meanwhile, unexpected experimental results from different fields of
physics could be explained qualitatively by means of this
phenomenon (see \cite{rotime} where some of them are sketched). 
The dynamical stabilization of the system is environmentally induced and
appears {\it only} in open quantum systems 
which are described by a non-Hermitian Hamilton operator due to the
embedding of the system into the environment of scattering wavefunctions.

It is the aim of the present paper to study in detail 
generic features of open quantum systems at
high level density where more than two states avoid crossing
and the ratio of the widths of the states to the energy differences 
between them is larger than 1. A realistic case of such a type is
considered first in nuclear physics \cite{klro}, then in
laser induced continuum structures in atoms \cite{marost}, and 
later in many other systems, see  the review \cite{top} and also 
\cite{rotime}. 
We restrict the study to four overlapping resonance states
and three different avoided  level crossings. 
In difference to the paper \cite{graefe} on the signatures of symmetric
Hamiltonians with three coalescing eigenfunctions, 
we are interested in the
behavior of the {\it eigenvalues} of a symmetric non-Hermitian operator
in the regime of overlapping resonances where the states {\it avoid crossing}. 
Here, the wavefunctions of the two states are mixed in a finite
parameter range around the critical  value $a_{\rm cr}$ of the 
avoided crossing. The mixing range shrinks to one point when
analytically continued up to the 
exceptional point by means of another parameter \cite{ro01}. At the
exceptional point, the
two eigenfunctions become linearly dependent from one another \cite{top}.

Our calculations are performed with real, complex as well as with
imaginary coupling coefficients of the states via the environment.  
The mixing of the wavefunctions caused by this coupling 
is simulated, according to the numerical results obtained in
\cite{ro01}, by assuming a Gaussian
distribution around the critical point of avoided crossing at the 
critical parameter value $a_{\rm cr}$.
At high level density, the Gaussian distributions related to
avoided crossings of different levels may overlap.
Our results  show that,
in the regime of overlapping  Gaussian distributions,
new effects arise from the imaginary part of the coupling term via 
the environment which are related to width bifurcation. They
are much more pronounced than the effects caused by the real part
of the coupling term which leads to level repulsion in energy. 
Furthermore, symmetries between the states are shown to play an 
important role in the avoided level crossing phenomenon.
This result hints at the relation of avoided level
crossings to  exceptional points. 

In section \ref{form}, the formalism used for the calculations, 
is formulated. Some results for two crossing levels are given in
section \ref{two} while those for four crossing levels 
can be found in section \ref{four}. 
The results are summarized and conclusions are drawn in the last
section.

\section{Formalism}
\label{form}

We consider an $N\times N$ matrix 
\begin{eqnarray}
{\cal H} = 
\left( \begin{array}{cccc}
\varepsilon_{1}  & \omega_{12} & \ldots &\omega_{1N}   \\
\omega_{21} & \varepsilon_{2}  &  \ldots & \omega_{2N}\\
\vdots     & \vdots &             \ddots&   \vdots \\
\omega_{N1} & \omega_{N2}       &    \ldots   &  \varepsilon_{N} \\
\end{array} \right)
\label{int1}
\end{eqnarray}
the diagonal elements of which are the $N$ complex eigenvalues 
$\varepsilon_{i} \equiv e_i - i/2~\gamma_i$
of a non-Hermitian operator. The $e_i$ and  $\gamma_i$ denote the 
energies and widths, respectively, of the $N$ states
without account of the interaction of the different states via an environment. 
This interaction is contained in the  values $\omega_{ik}$ which stand for the
coupling matrix elements $\langle \phi_i|{\cal H}|\phi_k\rangle$ 
of the states $i$ and $k$ via the environment, 
where $\phi_i$ is the wavefunction of the state $i$.
The corrections due to the coupling $\omega_{ii}$ of the state $i$ to the
environment (i.e. to the continuum of scattering wavefunctions
into which the system is embedded)
lead to  the selfenergy of the state $i$, see \cite{rotime}. 
In atomic physics, these corrections are  known as Lamb
shift.  With the only exception of figure \ref{fig7},
the $\omega_{ii}$ are assumed, in our model calculations,  
to be included into the $\varepsilon_i$. 
The $\omega_{ik}$ are complex, generally. The border case of purely  
imaginary  $\omega_{ik}$ corresponds to frozen internal
degrees of freedom \cite{top}.  For discrete states with
real energies $\varepsilon_i = e_i$, and the $\omega_{ik}$ are real
\cite{top,rotime}. 

The eigenvalues of $\ch$ will be denoted by $\ce_i 
\equiv E_i - i/2 ~\Gamma_i$ where  $E_i$ and $\Gamma_i$ stand for the
energy and width, respectively, of the eigenstate $i$. 
The  eigenfunctions $\Phi_i$ of ${\cal H}$ can be represented in the
set of basic wavefunctions $\phi_i$ of the unperturbed matrix
(corresponding to the case with vanishing coupling matrix elements
$\omega_{ij}$),
\begin{eqnarray}
\Phi_i=\sum_{j=1}^N b_{ij} \phi_j \; .
\label{int2}
\end{eqnarray}
The $b_{ij}$ are normalized  according to the biorthogonality
relations  of the wavefunctions $\{\Phi_i\}$ \cite{top}, 
\begin{eqnarray}
\langle \Phi_i^*|\Phi_j\rangle = \delta_{ij} \; .
\label{int3}
\end{eqnarray}
It follows 
\begin{eqnarray}
 \langle\Phi_i|\Phi_i\rangle = 
{\rm Re}~(\langle\Phi_i|\Phi_i\rangle) \;  ; \quad
 A_i \equiv \langle\Phi_i|\Phi_i\rangle \ge 1
\label{int4} 
\end{eqnarray}
and 
\begin{eqnarray}
\langle\Phi_i|\phi_{j\ne i}\rangle =
i ~{\rm Im}~(\langle\Phi_i|\phi_{j \ne i}\rangle) =
-\langle\Phi_{j \ne i}|\Phi_i\rangle \;  ;  \quad
|B_i^j|  \equiv 
|\langle \Phi_i | \Phi_{j \ne i}| ~\ge ~0  \; .
\label{int5}
\end{eqnarray}
The $\ce_i$ and $\Phi_i$ contain global features that are 
caused by many-body forces  induced by the coupling
$\omega_{ik}$ of the states via the environment, see \cite{rotime,jung}
and equations (7) and (8) in \cite{top}.

Some years ago, the case $N=2$ with $e_i=e_i(a)$, 
fixed real $\omega \equiv \omega_{12} = \omega_{21}$,
and different fixed values of $\gamma_i$, including $\gamma_i = 0$, 
is  studied  as a
function of the parameter $a$ in the neighborhood of avoided and true
crossings of the two levels \cite{ro01}.   In this case, the
two eigenvalues of ${\cal H}$ are
\begin{eqnarray}
\varepsilon_{i,j} \equiv e_{i,j} - \frac{i}{2} \gamma_{i,j} = 
 \frac{\varepsilon_1 + \varepsilon_2}{2} \pm Z; \quad \quad
Z \equiv \frac{1}{2} \sqrt{(\varepsilon_1 - \varepsilon_2)^2 + 4 \omega^2}
\; .
\label{int6}
\end{eqnarray}
According to this expression, two interacting discrete states (with
$\gamma_k = 0$) avoid always crossing since $\omega$ and 
$\varepsilon_1 - \varepsilon_2$ are real in this case \cite{top}. 
Resonance states with nonvanishing widths $\gamma_i$ 
repel each other in energy  according to Re$(Z) >0$
while the widths bifurcate according to   Im$(Z)  $.
The two states cross when $Z=0$. 

The results for the 
$N=2$ case  \cite{ro01} show further that the wavefunctions of the
two states $\Phi_1$ and $\Phi_2$ are mixed in a  finite range of the
parameter $a$ around the critical value $a_{\rm cr}$
 at which the two states
avoid crossing. This holds true not only for resonance states but also
for discrete states. Furthermore, a nonlinear source term appears in
the Schr\"odinger equation in the
neighborhood of an exceptional point and the critical point of an
avoided crossing, respectively. This source term causes irreversible
processes \cite{top,rotime}. 

In our calculations, the mixing coefficients $b_{ij}$ (see  equation
(\ref{int2})) of the wavefunctions of the two  states due to their avoided  
crossing are not calculated. We simulate the  fact  that  the two 
wavefunctions  are mixed in a finite parameter range around the
critical value of their avoid  crossing \cite{ro01} by assuming 
a  Gaussian distribution 
\begin{eqnarray}
\omega_{i\ne j} = \omega ~e^{-(e_i -e_j)^2}
\label{int7}
\end{eqnarray}
for the coupling coefficients.  
 
Calculations for  $N>2$  and for the case with complex $\omega$
are not performed up to now. Of special interest is the situation 
at high level density where the ranges of avoided crossings,
defined by (\ref{int7}),  of
different levels overlap.  Some generic results obtained  
with 2 and 4 resonance states will be presented in the following 
sections.

\section{${\bf \boldmath N=2}$  crossing levels}
\label{two}

\begin{figure}[ht]
\begin{center}
\includegraphics[width=7cm,height=3.4cm]{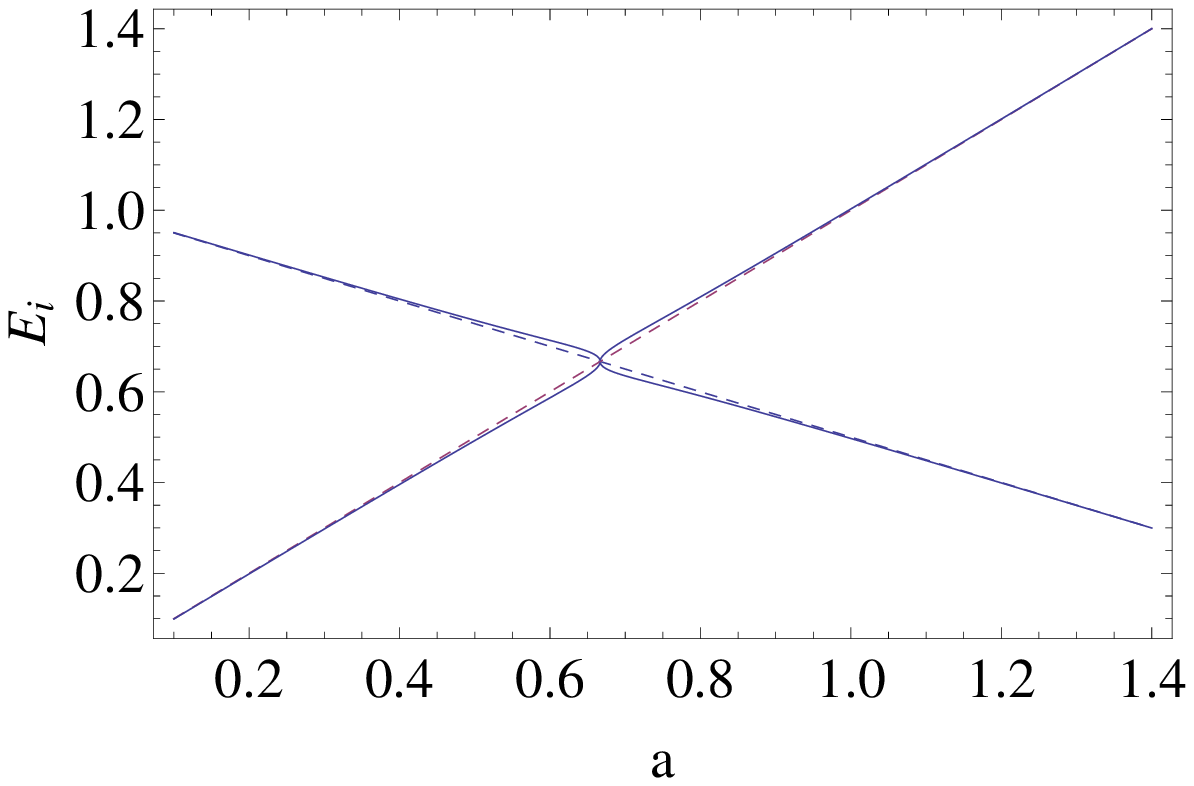}\\[.1cm] 
\includegraphics[width=7cm,height=3.4cm]{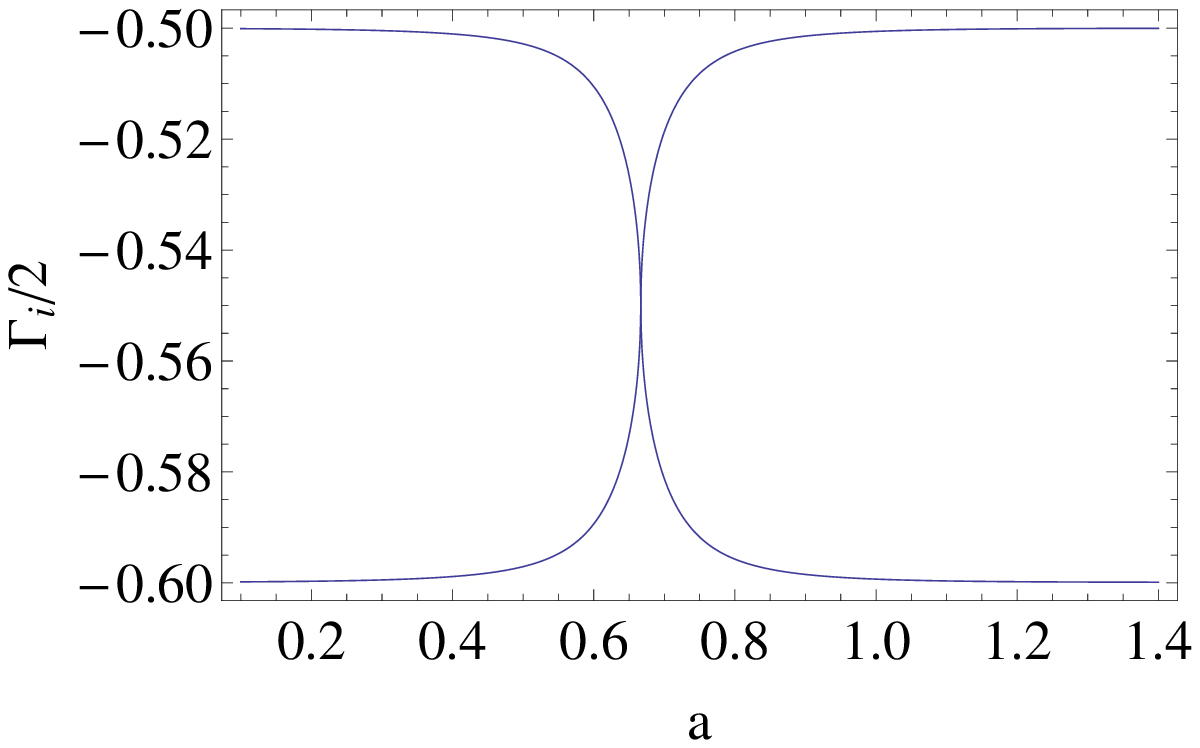}
\end{center}
\caption{\small 
The energies $E_i$ (top) and widths $\Gamma_i/2$ (bottom) for two
crossing states as a function of the parameter $a$ 
with $e_1 = 1-a/2; ~e_2= a$; ~~~$\gamma_1/2 = 0.5;  
~\gamma_2/2 = 0.5999$; ~~~$\omega = 0.05$. 
The dashed lines show the unperturbed  $e_i(a)$.
The exceptional point is at the crossing point of the 
energies $e_{1}$ and $e_2$. 
}
\label{fig1}
\end{figure}

\begin{figure}[ht]
\begin{center}
\includegraphics[width=7cm,height=3.4cm]{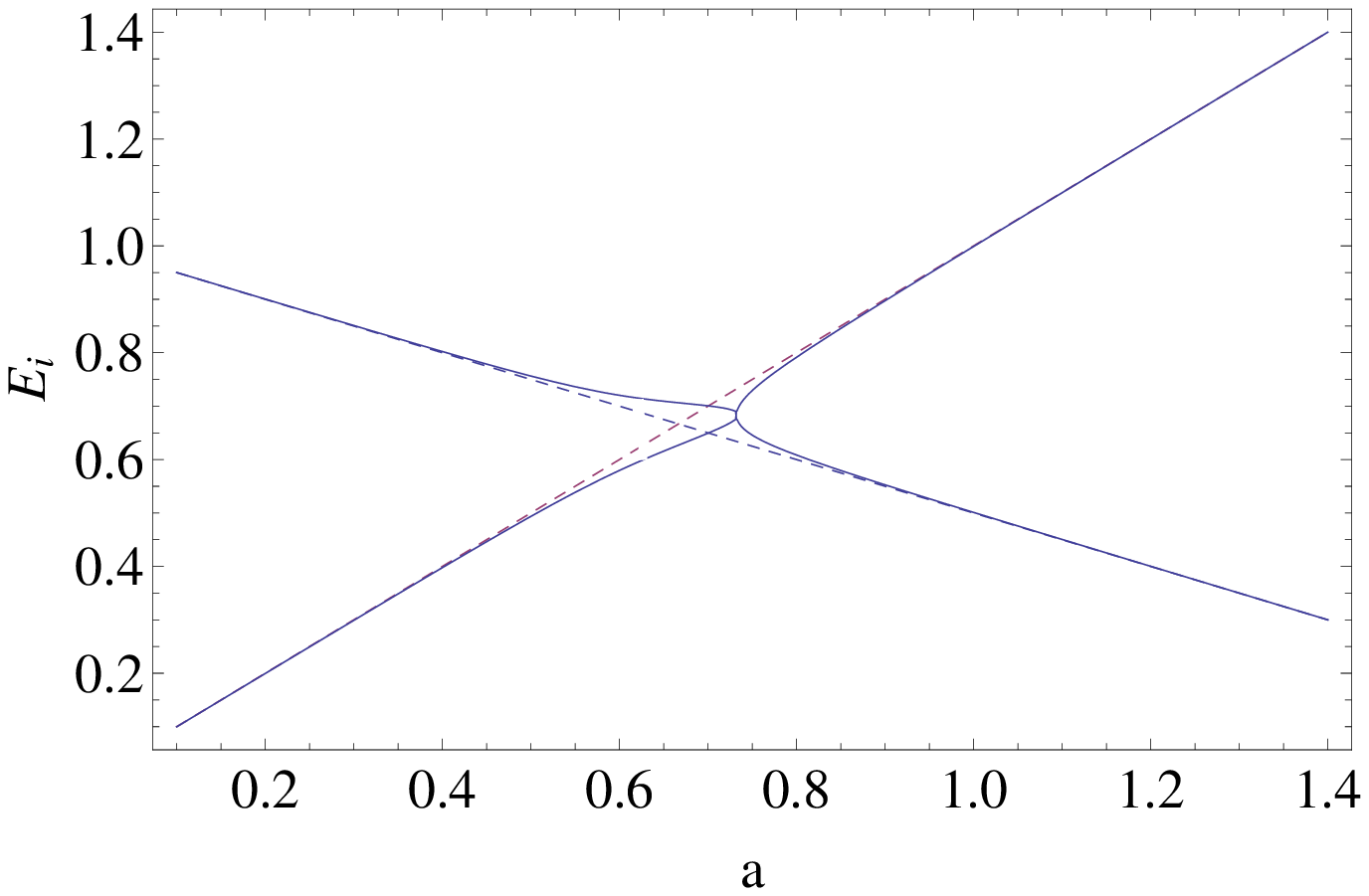}\\[.1cm] 
\includegraphics[width=7cm,height=3.4cm]{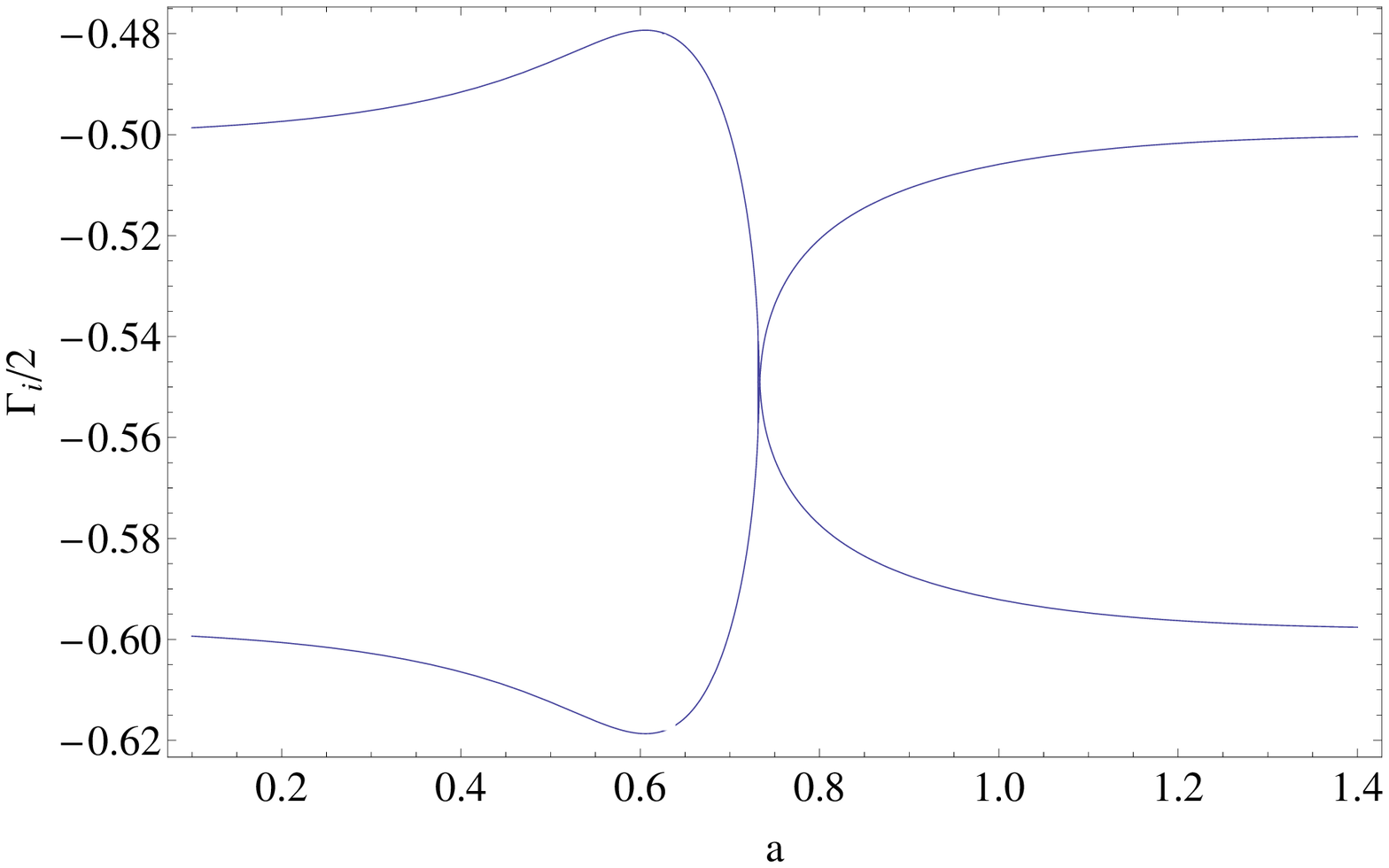}
\end{center}
\caption{\small 
The energies $E_i$ (top) and widths $\Gamma_i/2$ (bottom) for two
crossing states as a function of the parameter $a$  
with $e_1 = 1-a/2; ~e_2= a$; ~~~$\gamma_1/2 = 0.5;  
~\gamma_2/2 = 0.5980$; ~~~$\omega = 0.05 (1+i)$.
The dashed lines show the unperturbed  $e_i(a)$.
The exceptional point is shifted in $a$ relative to the crossing point of the 
energies $e_{1}$ and $e_2$. 
}
\label{fig2}
\end{figure}

First we study the two-level case most properties of which can be
found in the literature at different places. 
Here, we choose the matrix (\ref{int1}) with $e_i=e_i(a)$, 
different fixed values of $\gamma_i$ and  
fixed $\omega \equiv \omega_{12} = \omega_{21}$.
The functional dependence of the energies over the parameter $a$ is 
similar as in \cite{ro01}, but the $\omega$ may be complex.
We show the results   as a
function of the parameter $a$ in the neighborhood of avoided and true
crossings of the two levels.

In figure \ref{fig1}, the
energies $E_i$ and widths $\Gamma_i$ are shown for two crossing levels   
with the real coupling coefficient $\omega = 0.05$. At the critical
value  $a_{\rm cr}$ of the parameter $a$, the two energy
trajectories $e_i(a); ~i=1,2$ cross. Here, also 
the energy and width trajectories 
$E_i(a)$ and $\Gamma_i(a)$, respectively, may cross as shown in figure
\ref{fig1}. The crossing point of the $E_i(a)$ and $\Gamma_i(a)$
is an exceptional point. If one of the input values is
slightly different from the values used in figure \ref{fig1}, 
the $E_i(a)$ avoid crossing while the $\Gamma_i(a)$ cross
(or the $E_i(a)$ cross and the widths $\Gamma_i(a)$ bifurcate).  
In the first case,  
the two states are exchanged at the critical point  $a_{\rm cr}$
of the parameter $a$. 

The avoided crossing phenomenon of discrete states is known
in the literature for many years \cite{landau}.
As shown in \cite{ro01}, narrow resonance
states avoid crossing in energy in a similar manner as discrete
states. The only difference to the case with discrete states are
the nonvanishing widths $\Gamma_i(a)$ of the resonance states. 
The widths may cross at the critical parameter value $a = a_{\rm cr}$ if 
$\omega$ is real and the $E_i(a)$ avoid crossing, see 
equation (\ref{int6}).
 
In figure \ref{fig2}, the situation with two crossing levels but
complex coupling $\omega$ is shown. Also in this case an exceptional
point can be found,  though for a value of $\gamma_2$ being different
from that in figure \ref{fig1} due to the other value of $\omega$.  
In figure \ref{fig2}, the position of the exceptional point
is shifted in relation to the crossing point of
the $e_i(a)$ trajectories. Furthermore, the difference $|\Gamma_1 - \Gamma_2|$ 
blows up around the crossing point of the $e_i$. However, at large distances
of the parameter $a$ from the critical value, the widths of the two
states approach quickly the  values $\gamma_1$ and $\gamma_2$, respectively.    
If one of the input values is slightly changed, the two eigenvalue
trajectories avoid crossing in energy similar as in the case with real coupling
$\omega$ (figure \ref{fig1}) while the $\Gamma_i(a)$ cross freely,
or the $E_i(a)$ cross and the $\Gamma_i(a)$ bifurcate.
In the first case, the two states are exchanged 
at the critical parameter value at which the two levels avoid crossing.

\begin{figure}[ht]
\begin{center}
\includegraphics[width=7cm,height=3.4cm]{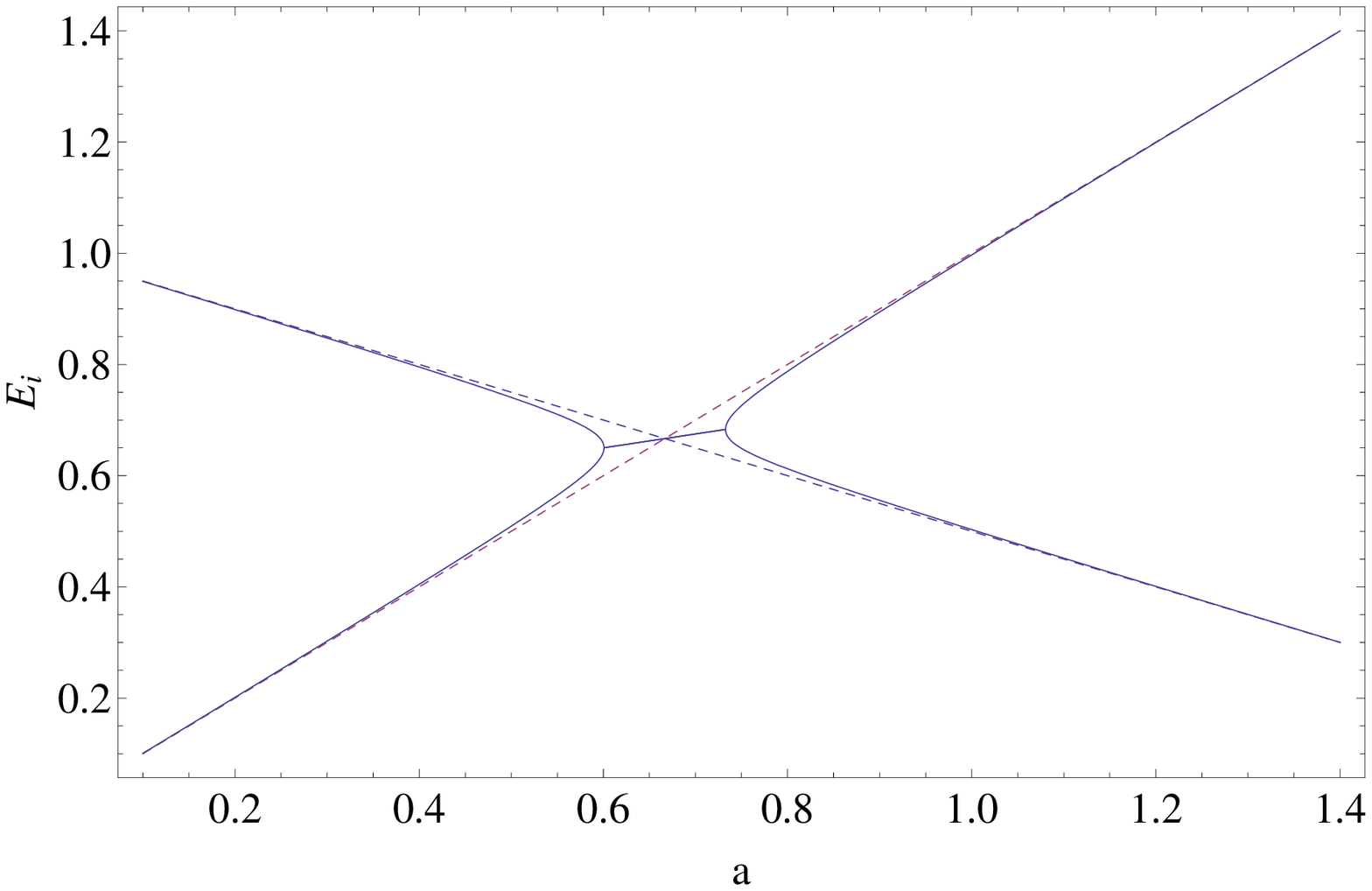} \\[.1cm]
\includegraphics[width=7cm,height=3.4cm]{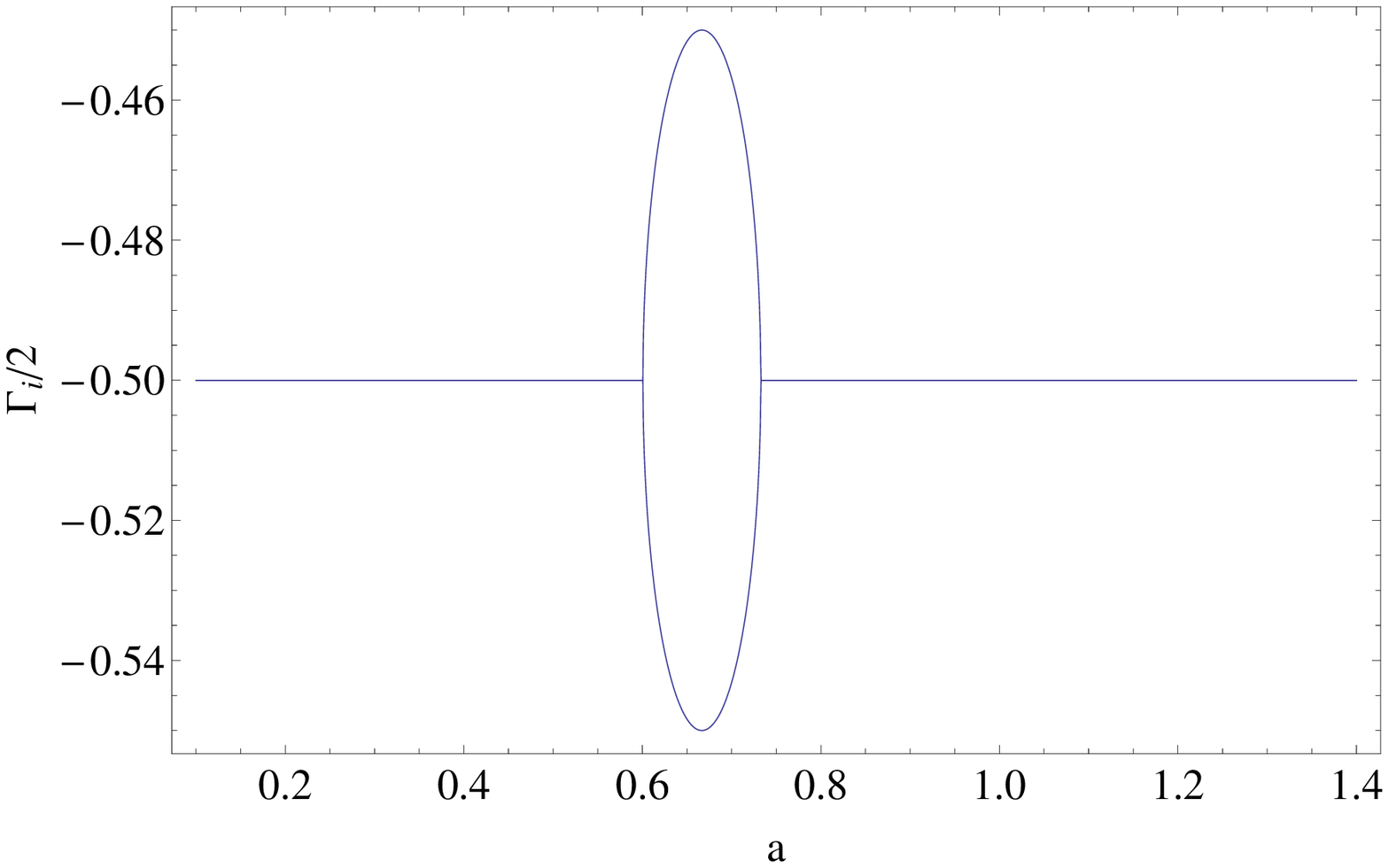}
\end{center}
\caption{\small 
The energies $E_i$ (top) and widths $\Gamma_i/2$ (bottom) for two
crossing states as a function of the parameter $a$ 
with $e_1 = 1-a/2; ~e_2= a$; ~~~$\gamma_1/2 =   
\gamma_2/2 = 0.5$; ~~~$\omega = 0.05 i$. 
The dashed lines show the unperturbed  $e_i(a)$.
The widths bifurcate in the parameter range where $E_1 = E_2$.
}
\label{fig3}
\end{figure}

The blowing up of the difference $|\Gamma_1 - \Gamma_2|$ around $a_{\rm  cr}$
seen in figure \ref{fig2}, is nothing but  width bifurcation according
to Im$(Z)$, equation (\ref{int6}). This can be seen very clearly in
figure \ref{fig3} calculated with imaginary $\omega$. In a
relatively small parameter range around the crossing point of the $e_1$
and $e_2$ trajectories, we have
$E_1 = E_2$ while the difference  $|\Gamma_1 - \Gamma_2|$ is large.
Beyond this parameter range, quickly $E_i \to e_j$ and  
$\Gamma_i \to \gamma_j$. The width bifurcation is caused by Im$(\omega)$.
It plays an important role for resonance states that cross at
high level density since, in difference to the avoided crossing of discrete
states, the coupling of resonance states via the environment is
complex in this case \cite{top}. 

Figures \ref{fig1} to \ref{fig3} show the characteristic features of
the dynamics of an open quantum system in the neighborhood of an
exceptional point: the two crossing states are exchanged at the
critical value of the control parameter including their population
and, furthermore, the widths bifurcate due to Im{$(\omega)$}. Population 
transfer occurs in realistic systems. A few examples are 
the high-order harmonic
generation in a driven two-level atom \cite{carla},
the ultrafast stimulated Raman parallel adiabatic passage by shaped
pulses \cite{raman} and 
the molecular vibrational cooling by adiabatic population transfer 
from excited to ground vibrational states \cite{atabek}. 
Width bifurcation becomes important at high level density as will be
shown in the next section.

\section{${\bf \boldmath N=4}$ crossing levels}
\label{four}

\begin{figure}[ht]
\begin{center}
\includegraphics[width=7cm,height=3.4cm]{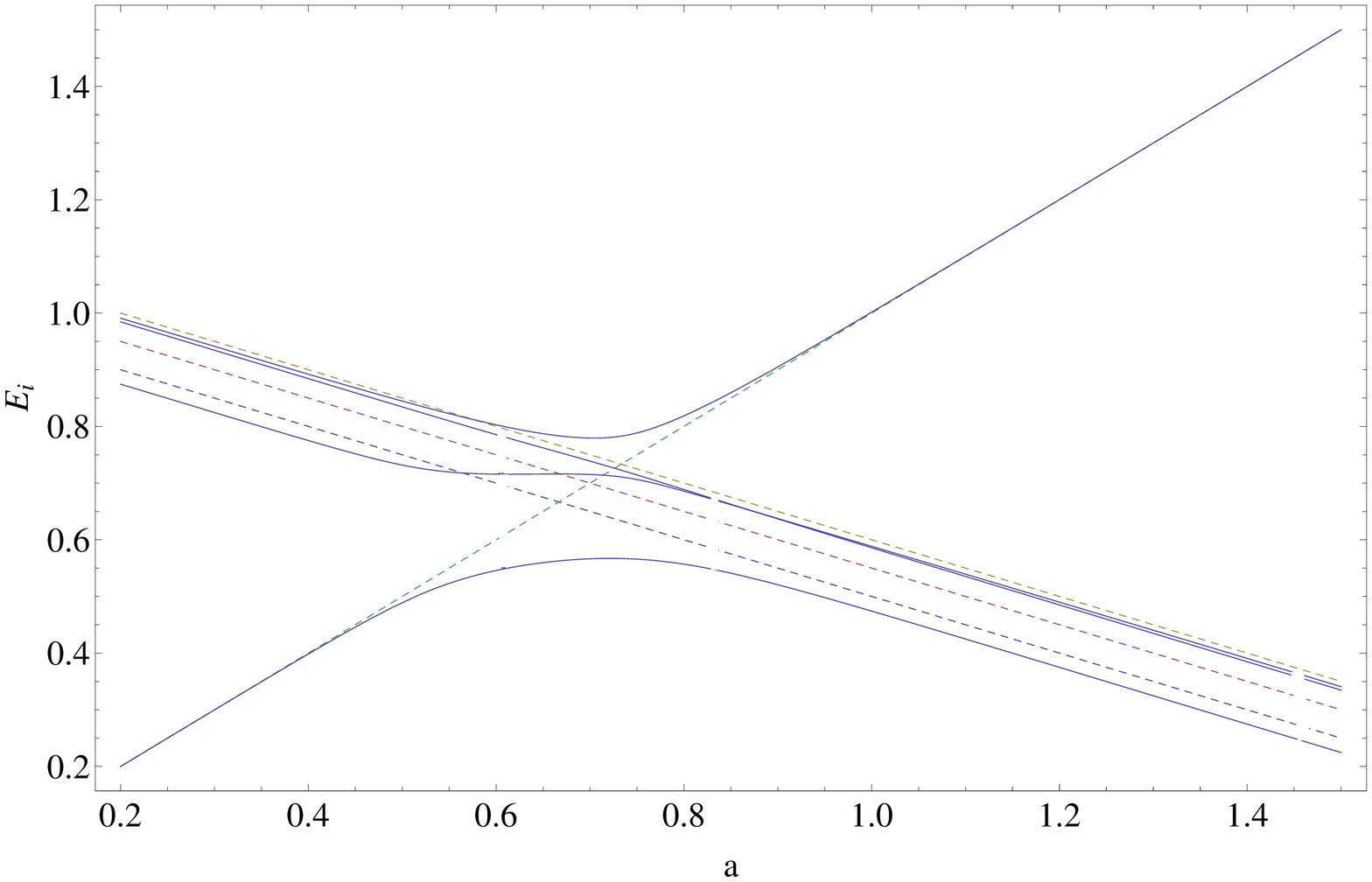} \\[.1cm]
\includegraphics[width=7cm,height=3.4cm]{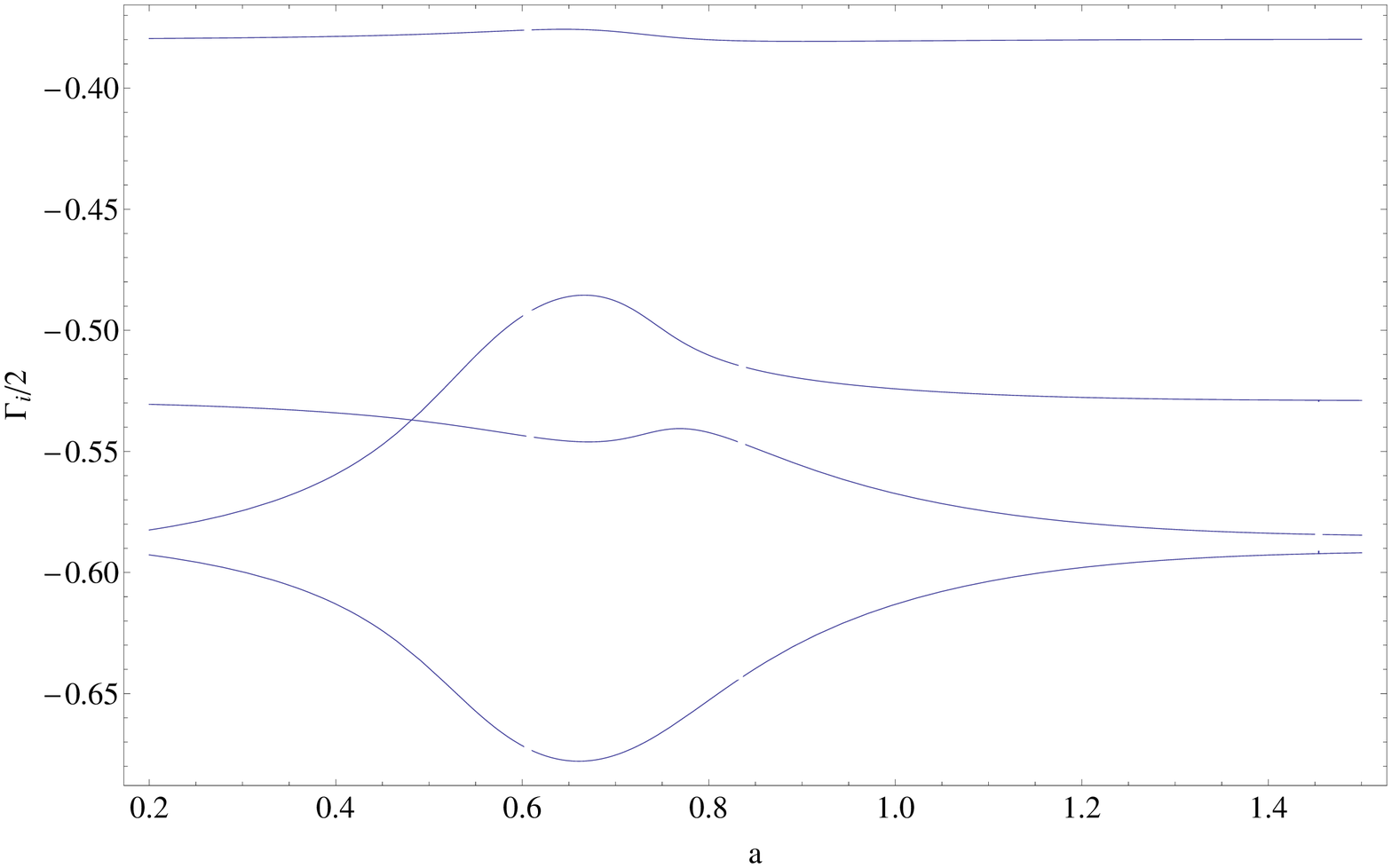}
\end{center}
\caption{\small 
The energies $E_i$ (top) and widths $\Gamma_i/2$ (bottom) for one
state crossing three other states as a function of the parameter
$a$ with $e_1 = 1-a/2; ~e_2 = 1.05 - a/2; ~ e_3 = 1.1 - a/2; ~ e_4 
= a$; ~~~$\gamma_1/2 = 0.5; ~\gamma_2/2 = 0.4; ~\gamma_3/2 = 0.6; 
~\gamma_4/2 =0.58523$;
~~~$\omega = 0.05~(1+i) $.
The dashed lines show the unperturbed  $e_i(a)$.
The widths $\Gamma_i$ of the three overlapping states 1, 3 and 4 
bifurcate (or cross freely) in the parameter range of intersection
while energy $E_2 \approx e_2$ and
width  $\Gamma_2 \approx \gamma_2$ remain almost unaffected.
}
\label{fig4}
\end{figure}

\begin{figure}[ht]
\begin{center}
\includegraphics[width=7cm,height=3.4cm]{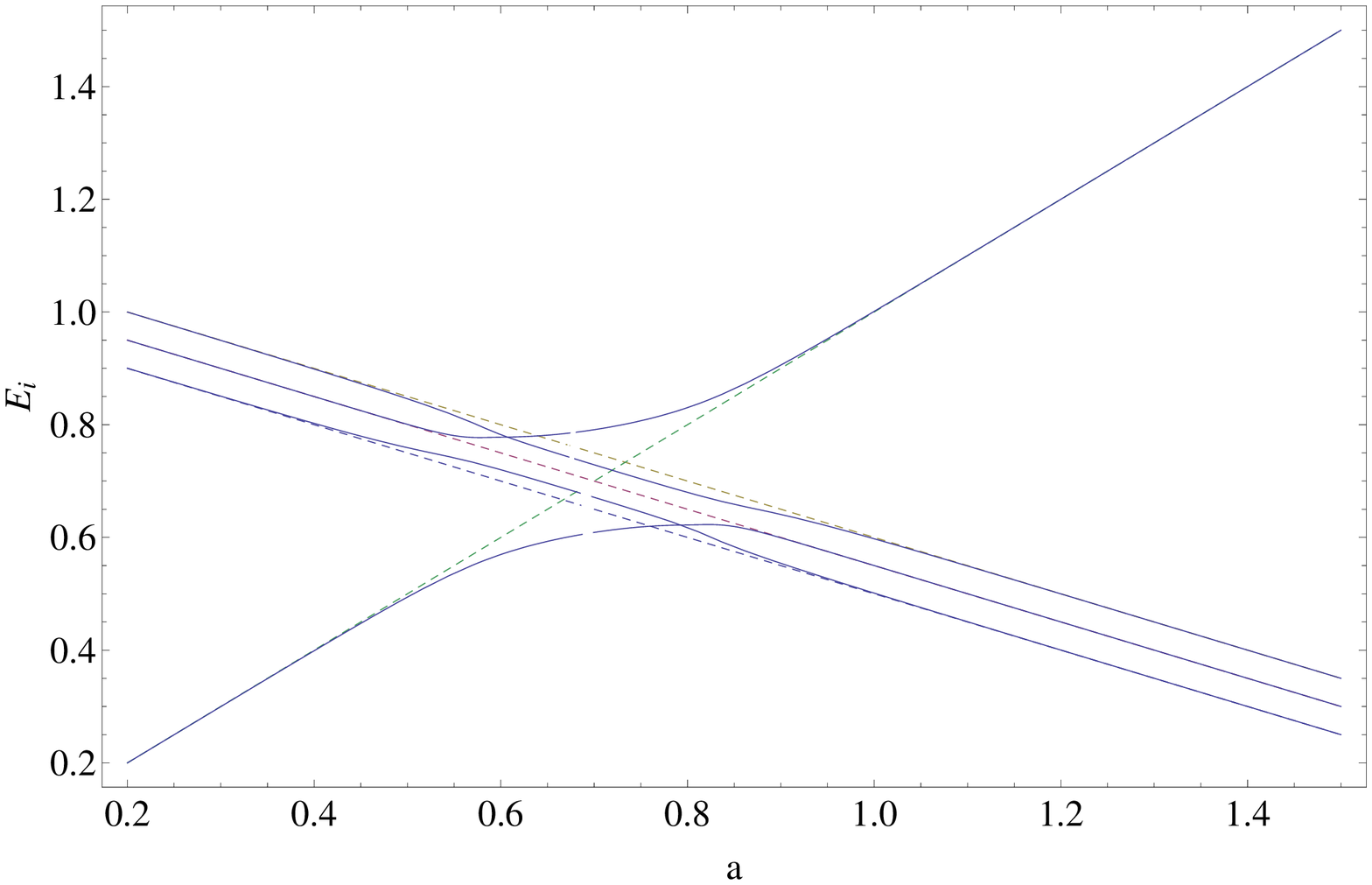} \\[.1cm] 
\includegraphics[width=7cm,height=3.4cm]{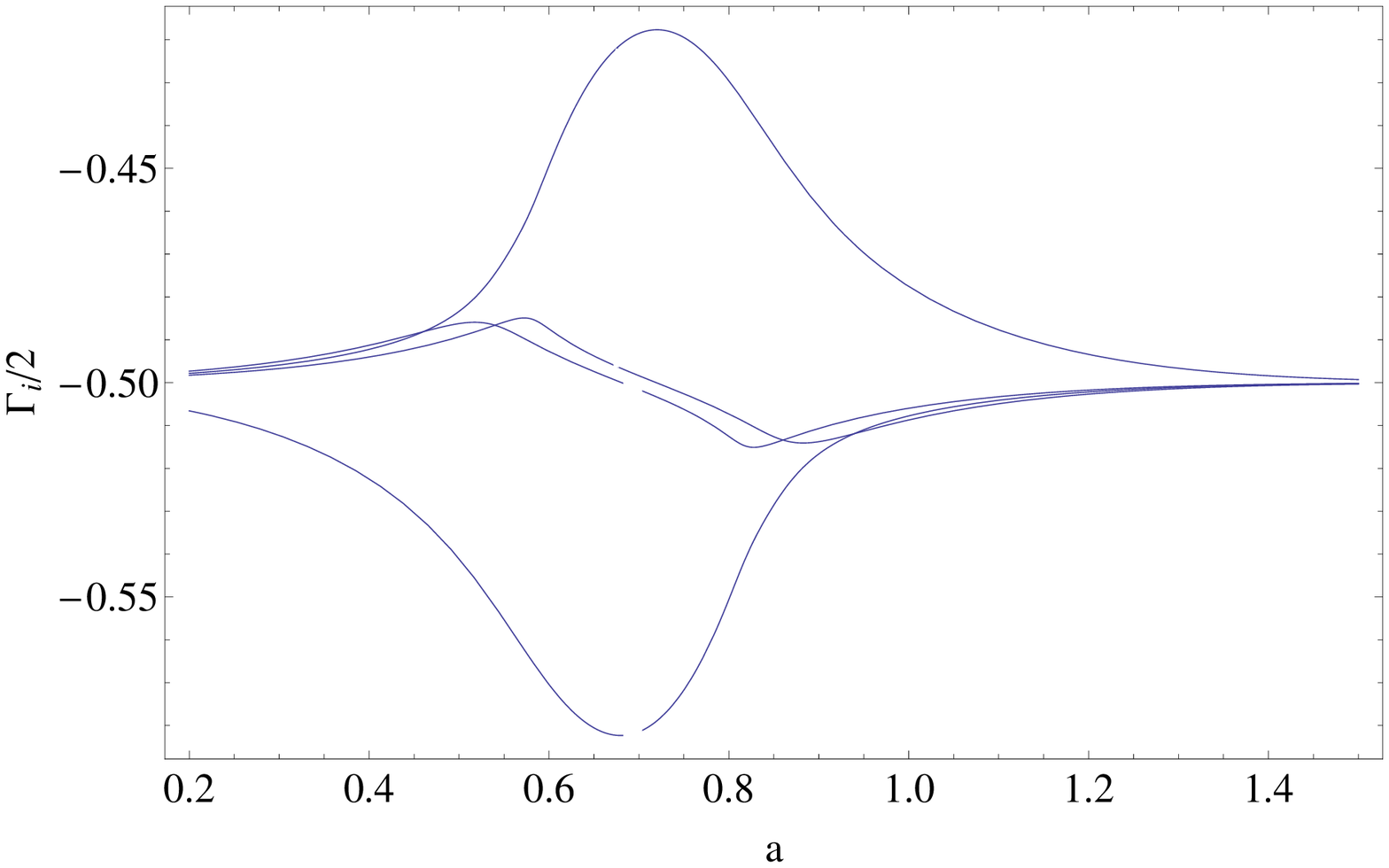}
\end{center}
\caption{\small 
The same as figure \ref{fig4} but 
~$\gamma_1/2 =  \gamma_2/2 = \gamma_3/2 = \gamma_4/2 = 0.5$.
\hspace{.1cm} $\omega = 0.05~(1+i)$.
All states participate in the interaction scenario. Far from the 
critical parameter range, all $\Gamma_i$ approach the value 0.5.
}
\label{fig5}
\end{figure}

\begin{figure}[ht]
\begin{center}
\includegraphics[width=7cm,height=3.4cm]{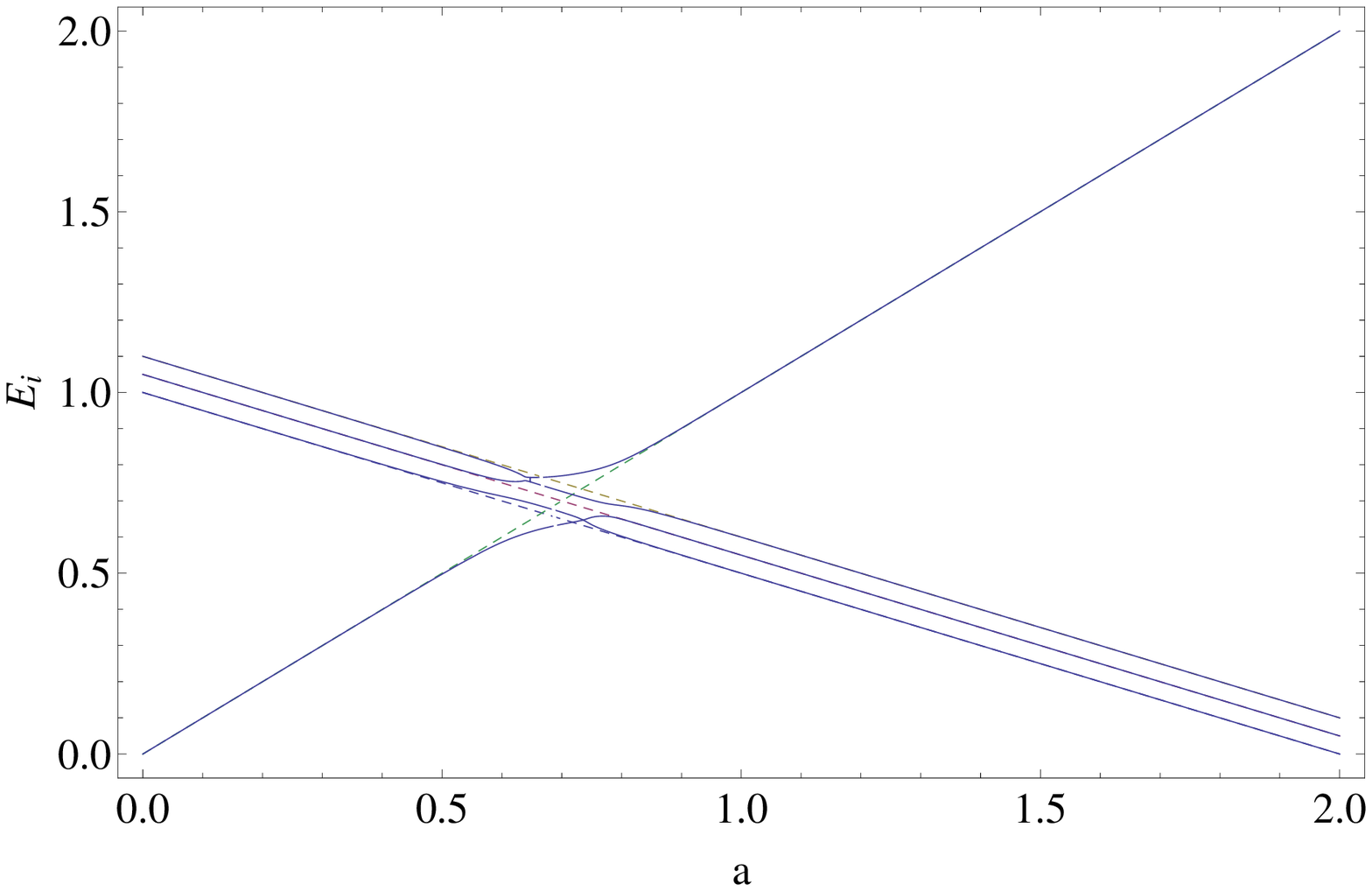} \\[.1cm]
\includegraphics[width=7cm,height=3.4cm]{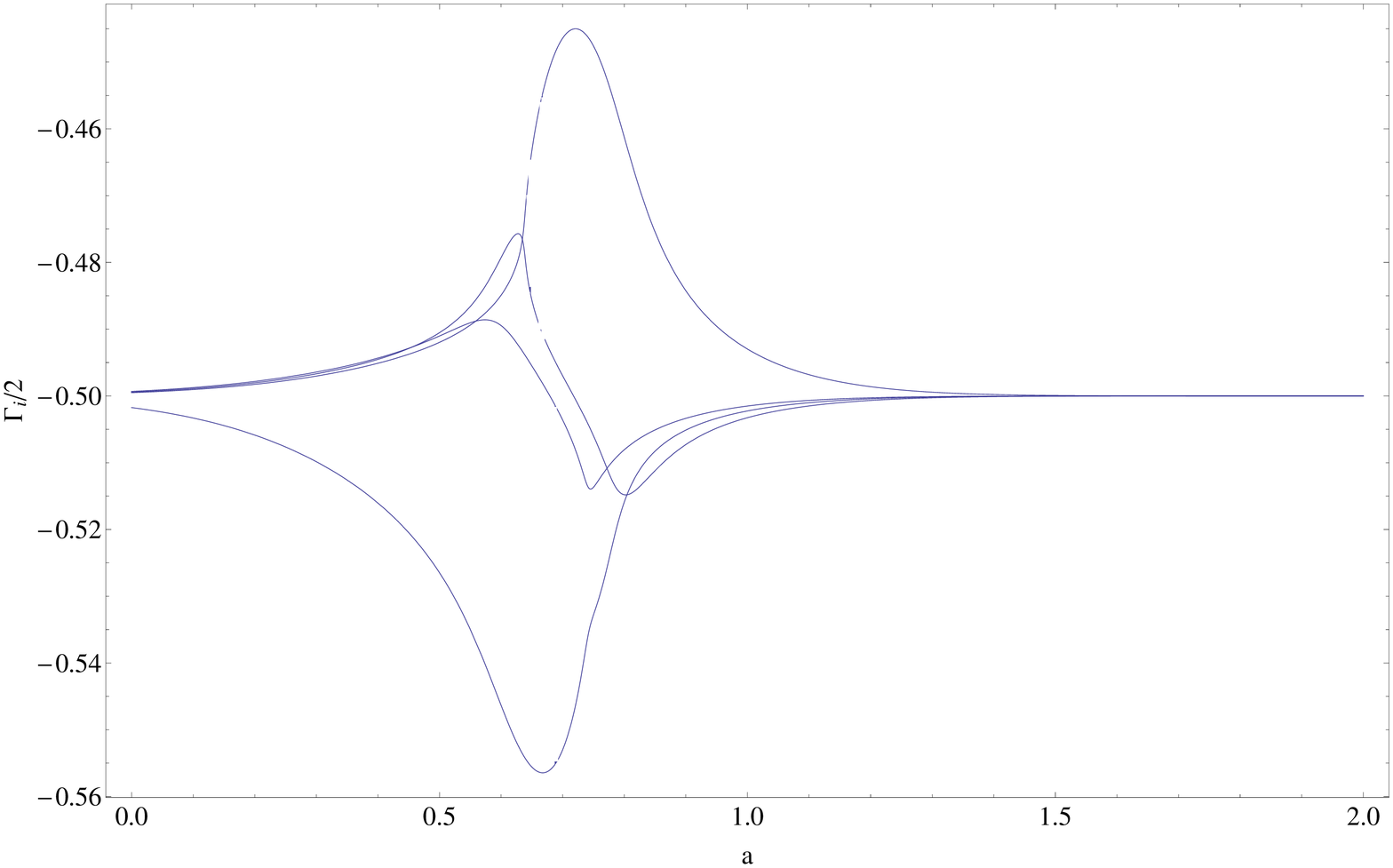}
\end{center}
\caption{\small 
The same as figure \ref{fig5} but
$\omega_{i4}= \omega_{4i} = \omega ~e_i(a) ~e^{-(e_i -e_4)^2}$; 
~~~$\omega_{i ~j\ne 4} = \omega_{j\ne 4 ~i} =0; ~~~\omega_{44} = 0 $.
\hspace{.1cm} $\omega = 0.05(1+i)$. 
The widths bifurcate less than in figure \ref{fig5}.
All states participate in the intersection scenario.
Far from  the critical parameter range 
all $\Gamma_i$ approach the value 0.5. 
}
\label{fig6}
\end{figure}

\begin{figure}[ht]
\begin{center}
\includegraphics[width=7cm,height=3.4cm]{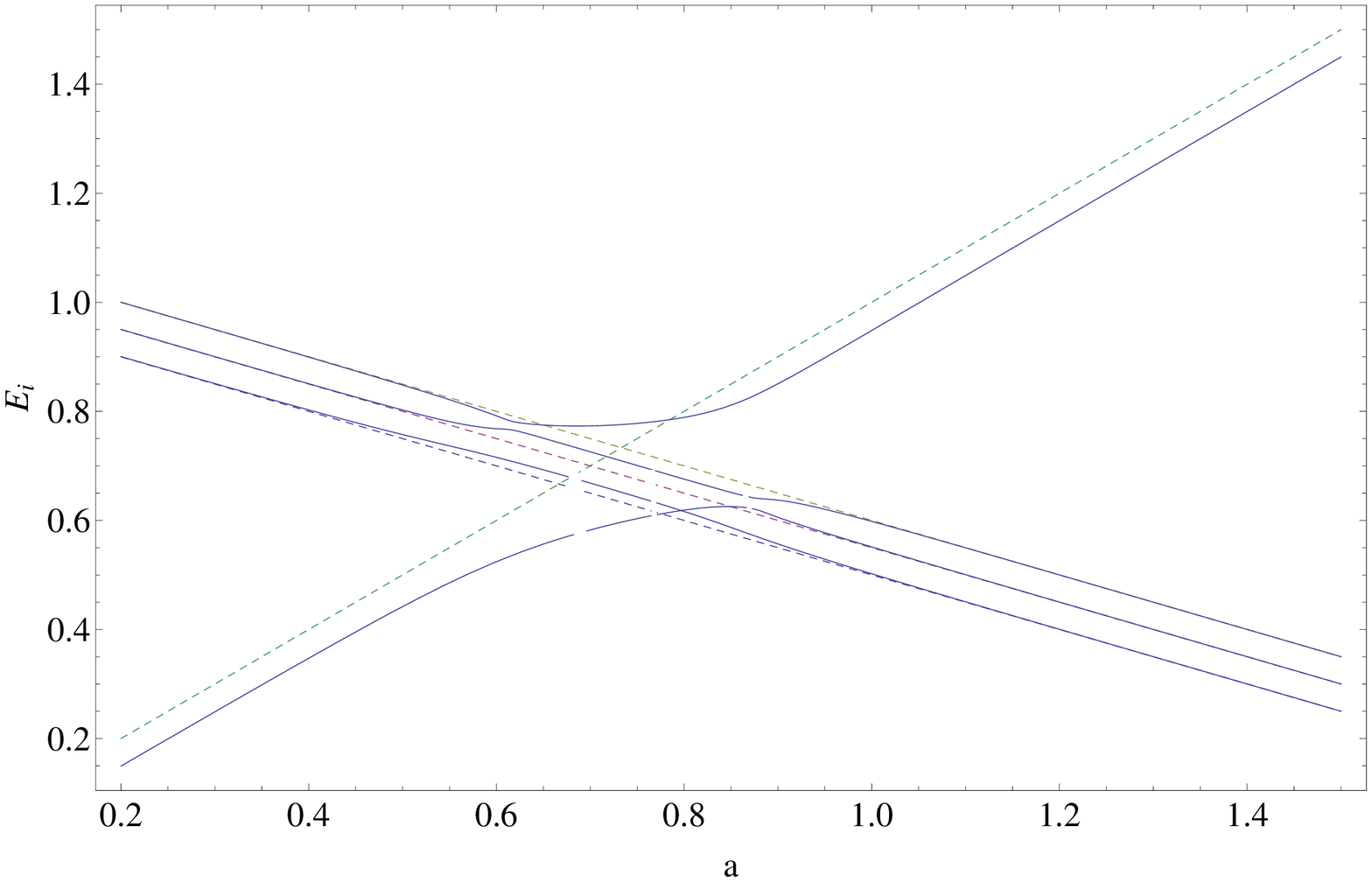} \\[.1cm] 
\includegraphics[width=7cm,height=3.4cm]{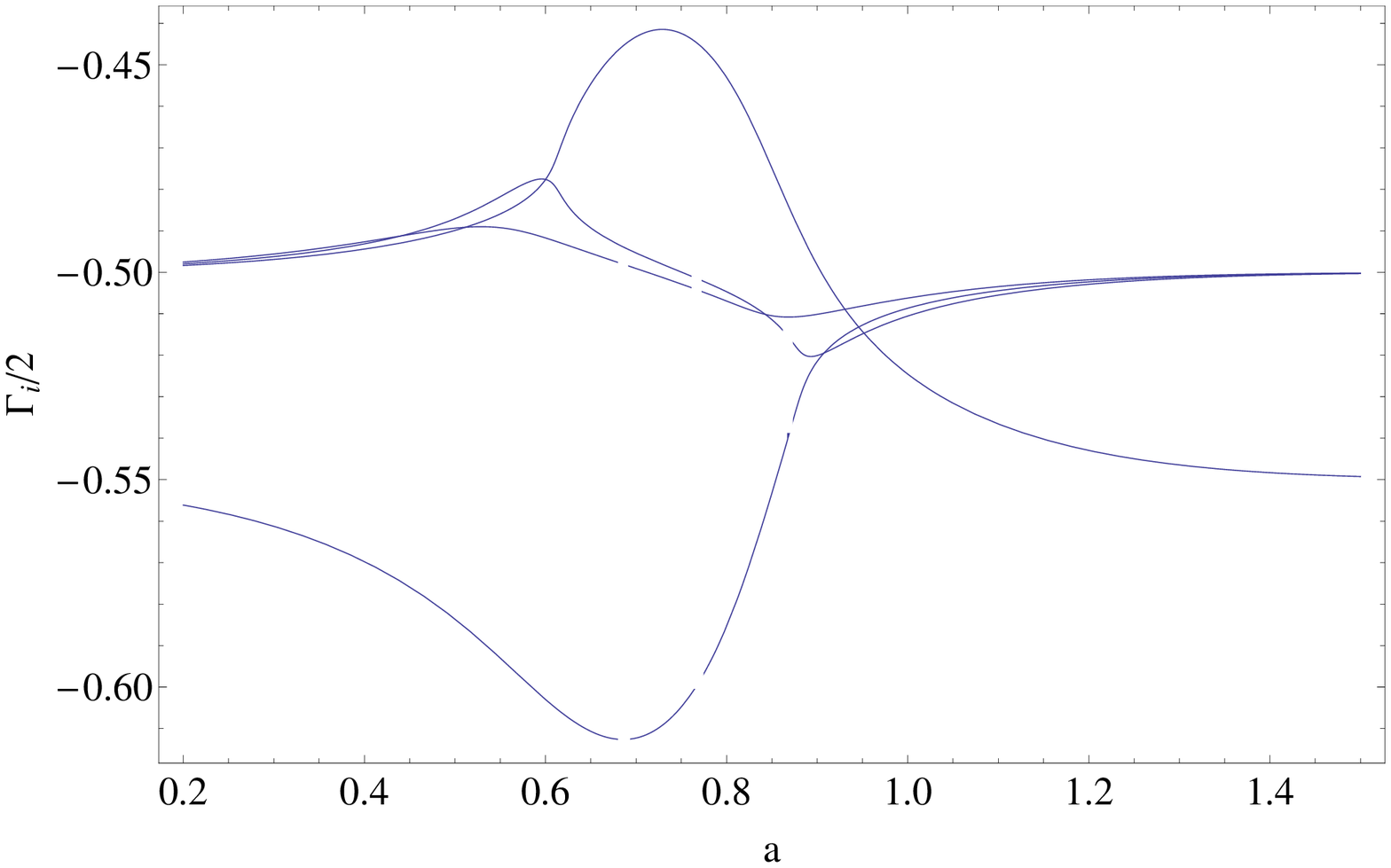}
\end{center}
\caption{\small 
The same as figure \ref{fig6} but 
$\omega_{44} = \omega = 0.05(1+i)$.
The widths $\Gamma_i$  bifurcate
in the parameter range of intersection.
The selfenergy term $\omega_{44}\ne 0$ causes  shift in energy and 
width far from the critical parameter range.
}
\label{fig7}
\end{figure}

\begin{figure}[ht]
\begin{center}
\includegraphics[width=7cm,height=3.4cm]{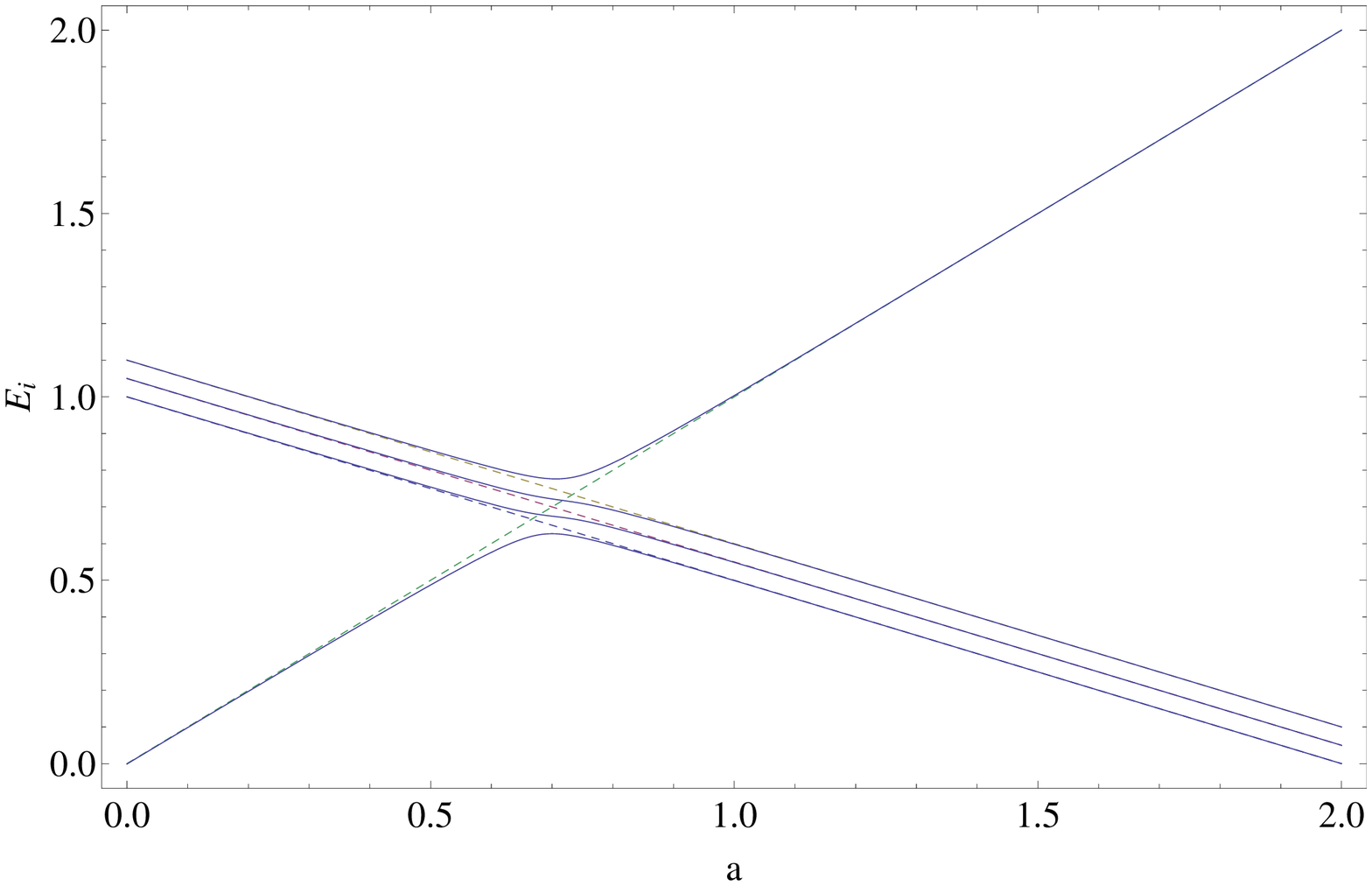}\\[.1cm] 
\includegraphics[width=7cm,height=3.4cm]{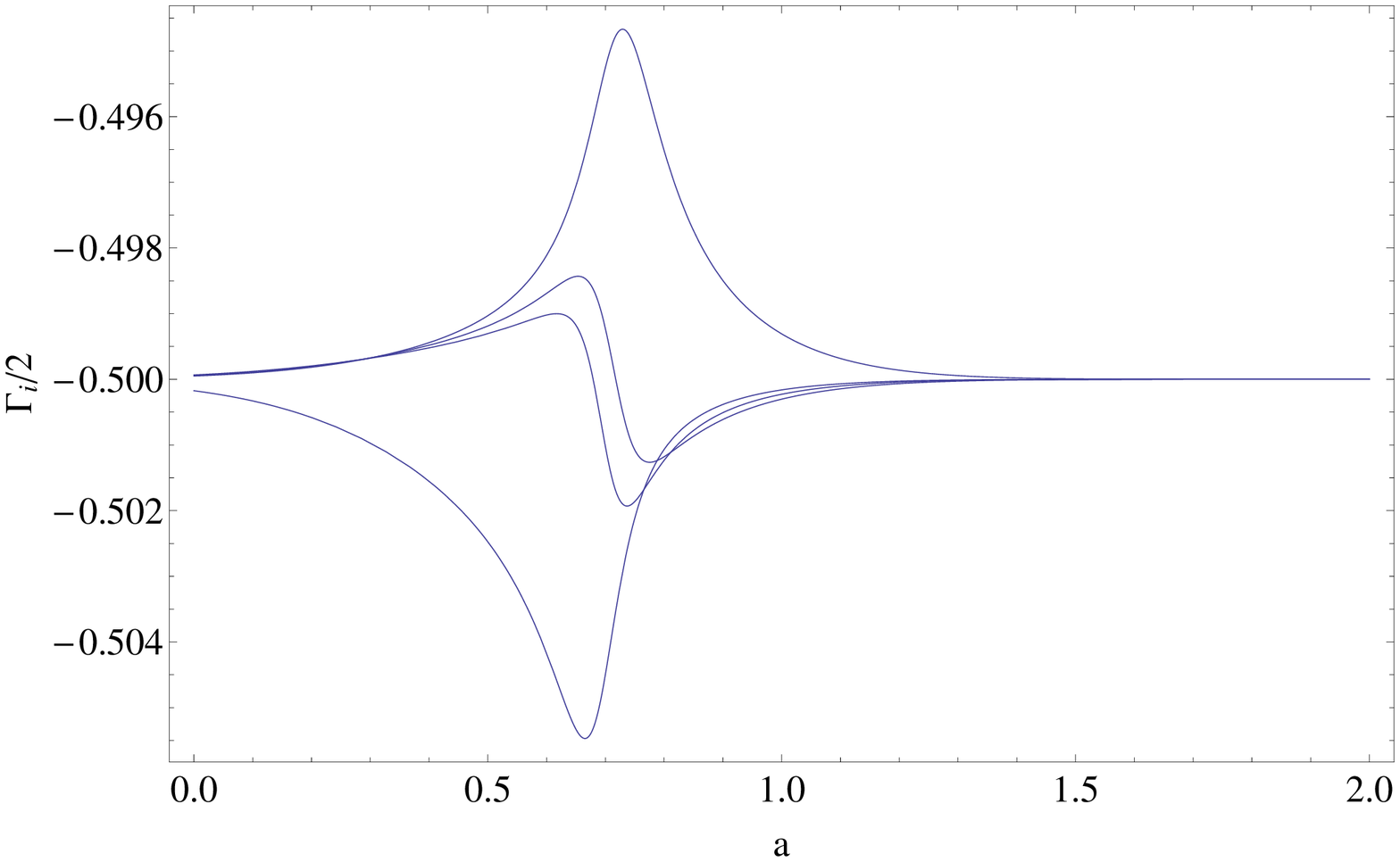}
\end{center}
\caption{\small 
The same as figure \ref{fig6} but $\omega = 0.05~(1+i/10)$. 
All states participate in the avoided level crossings.
Far from  the critical parameter range 
all $\Gamma_i$ approach the value 0.5.
}
\label{fig8}
\end{figure}

\begin{figure}[ht]
\begin{center}
\includegraphics[width=7cm,height=3.4cm]{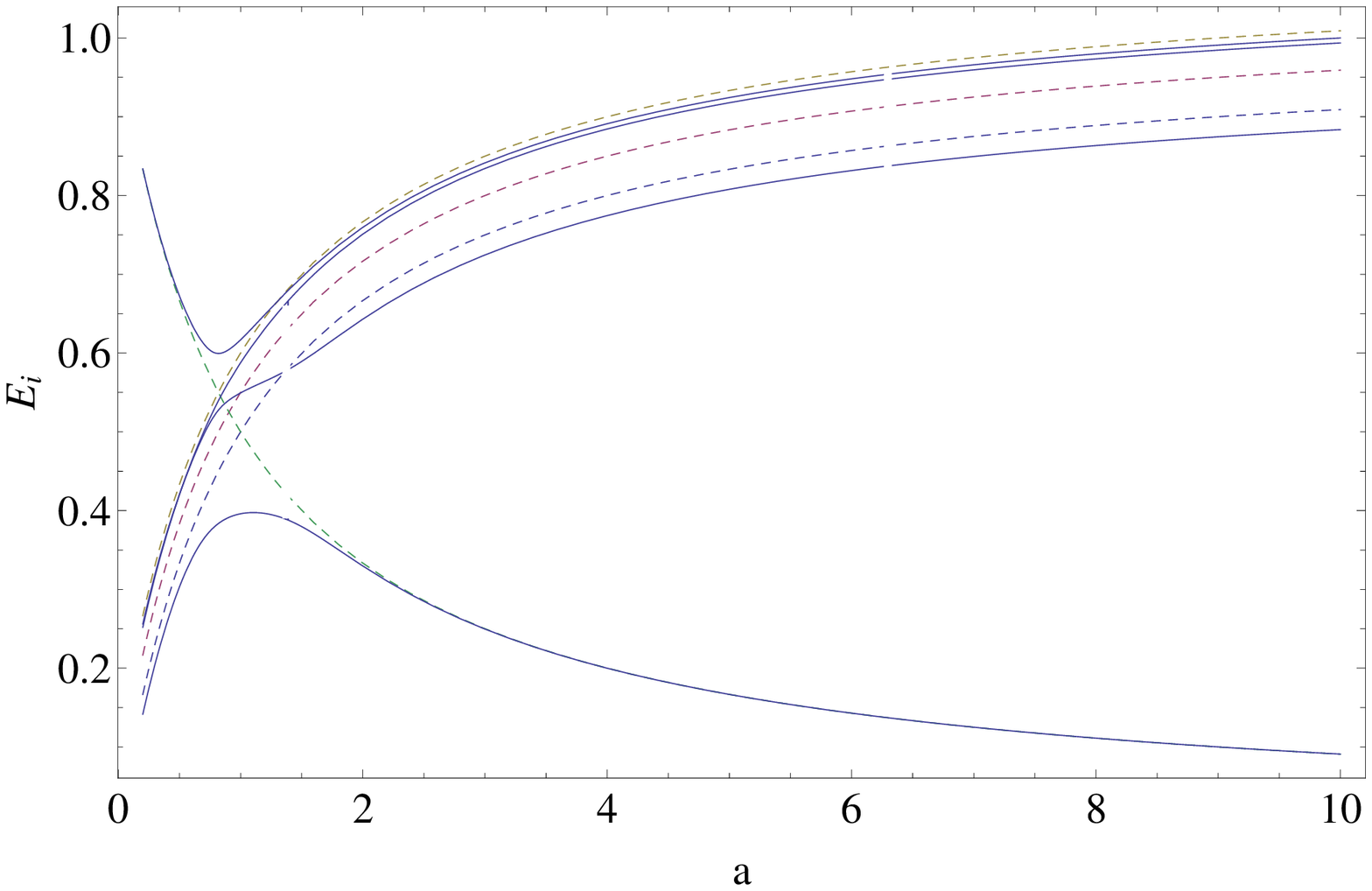} \\
\includegraphics[width=7cm,height=3.4cm]{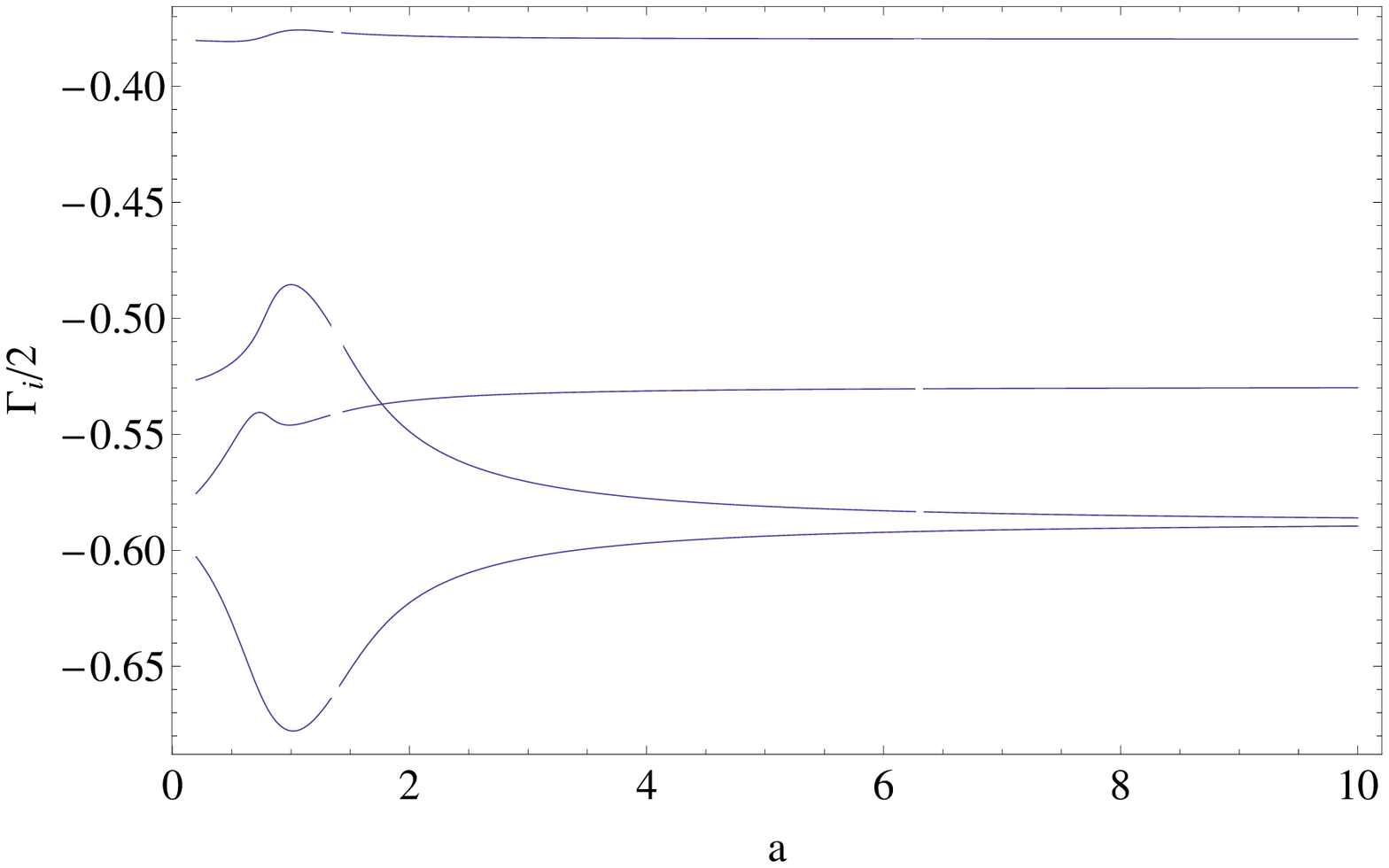}
\end{center}
\caption{\small 
The energies $E_i$ (top) and widths $\Gamma_i/2$ (bottom) for four
crossing states  as a function of the parameter $a$
with $e_1 = 1-\frac{1}{a+1}; ~e_2= 1.05 - \frac{1}{a+1}; ~e_3=1.1
-\frac{1}{a+1}; ~e_4= \frac{1}{a+1}$;
~~~$\gamma_1/2 = 0.5;  ~\gamma_2/2 = 0.4; ~\gamma_3/2 = 0.6;
~\gamma_4/2 = 0.58523$; ~~~$\omega = 0.05~(1+i)$.
The dashed lines show the unperturbed  $e_i(a)$.
The widths $\Gamma_1,  ~\Gamma_3, ~\Gamma_4$ 
bifurcate in the critical parameter range while the width
$\Gamma_2 \approx \gamma_2$ and the energy $E_2 \approx e_2$ remain 
almost unchanged (as in figure \ref{fig4}).
}
\label{fig9}
\end{figure}

Some results of calculations with one resonance state crossing three
other ones are shown in figures \ref{fig4} up to \ref{fig9}. 
Mostly we use a linear functional dependence of the energies
$e_i(a)$ on the control parameter $a$.
Figure \ref{fig4} shows the role,  the symmetry in the 
overlapping of resonances plays in the critical region 
of the parameter $a$.
Energy $e_2$ and width $\gamma_2$ of the state 2 are 
fully symmetric with correspondence to the two states 1 and 3. 
The avoided crossings of the state 2 with the
states 1 and 3, respectively, are disturbed therefore
completely symmetrically by the states 3 and 1, respectively. Due to
this fact, the state 2 does (almost) not take place in the avoided level
crossing phenomenon  but appears as an 'observer'. 
Both, the energy trajectories $E_2$ and the width trajectories
$\Gamma_2$ follow the trajectories $e_2$ and $\gamma_2$ everywhere
including the parameter values $a$ around  $a_{\rm cr}$. 
Such an effect is known from a realistic case with 3 crossing levels
in a quantum dot \cite{rosad}.

The crossings and avoided crossings in energy together with the
corresponding bifurcations and crossings in width 
of the other states 1, 3 and 4 can be seen very
clearly in figure \ref{fig4}. Far from the crossing region, all 
eigenvalue trajectories  $\Gamma_i; ~i=1, 3, 4$ approach  those of the
$\gamma_{i \ne j}$.  An exchange of the states takes place in the
critical region around $a_{\rm cr}$ with the only exception of state 2.

The numerical symmetry of state 2 relative to the states 1 and 3 is somewhat 
disturbed when the widths $\gamma_i$ of all the states are equal to one another,
see figure \ref{fig5}. In this case, all four states are affected by the
crossings. In the critical region, the avoided level crossings in
energy and free crossings in width as well as the free crossings in
energy and bifurcations in width can be seen.
Far from the critical region, all eigenvalue trajectories
$E_i$ and $\Gamma_i$ approach the trajectories $e_j$ and 
$\gamma_{j}$, respectively. The small shift in energy seen in figure \ref{fig4}
does not appear in figure \ref{fig5} due to the same values of all
$\gamma_i$ in the last case. This shift is an indicator of the 
residual influence of
the intersection onto the eigenvalues of $\ch$ far from the
critical region with avoided level crossings. In other words, it is a
signature of the different couplings of the states to the environment 
in the regime of overlapping.

In difference to the results shown in figures \ref{fig4} and
\ref{fig5}, in those of
figure \ref{fig6} the states 1, 2 and 3 do not interact via the
environment: $\omega_{12}=\omega_{13}=\omega_{23} =0$. Their interaction
with state 4 via the environment is $\omega_{i4}=\omega_{4i} = \omega
e_i(a) e^{-(e_i-e_4)^2}$. The results are very similar to those shown in
figure \ref{fig5}, except for the fact that the overall interaction is
smaller. 

Figure \ref{fig7} shows
the influence of the selfenergy term $\omega_{44} \ne 0$. It causes 
some shift in all $E_i$ and all $\Gamma_i$ what can be seen best far
from the critical region.  

The relative influence of the imaginary part of the coupling strength
$\omega$ in the region around $a_{\rm cr}$ can be seen in figure
\ref{fig8}. The calculations are performed with the same values as
in figure \ref{fig6} but $\omega = 0.05(1+i/10)$ instead of $\omega =
0.05(1+i)$. The width bifurcation in figure \ref{fig8} is smaller than
that in figure \ref{fig6} due to the smaller value of Im$(\omega)$.

In figure \ref{fig9}, we show the results for a completely other 
distribution of the levels  in energy.
The functional dependence of the energies is chosen according to
a Coulomb-like potential.
All the features discussed in figures \ref{fig4} to \ref{fig8} 
can be seen also with this distribution. Even the influence of the symmetry of
the states 1 and 3 in relation to the state 2 causes the state 2
to be an 'observer', in the same manner as discussed in figure \ref{fig4}.
Further results obtained  in other calculations 
with the $e_i$ distribution of figure \ref{fig9} 
are  not shown in the present paper.  

Similar results are obtained also for the case that two 
resonance states cross two
other ones according to, e.g., $e_1 = 1-a/2; ~~ e_2=1.05 - a/2; 
~~ e_3=0.05+a; ~~ e_4 =a$, and different values for the $\gamma_i$.
Numerical results have shown furthermore that the Gaussian distribution 
(\ref{int7}) used in the calculations, ensures the overlapping of the 
different avoided crossings, i.e. the overlapping
of the parameter ranges in which the wavefunctions of the two crossing
states  are mixed.  Results with a broader Gaussian distribution 
are almost the same as those shown in the figures of the present paper.   

The features caused by the true and avoided crossings shown in the
figures \ref{fig4} to \ref{fig9} are generic. In any case, 
the influence of the level crossings occurring in a relatively small 
region  around the critical value  $a_{\rm cr}$, 
onto the four eigenvalues of $\ch$ is, generally, large.
Only at large distances from the critical region, the widths 
$\Gamma_i$ of the states approach the values $\gamma_{j}$ with 
$j \ne i$ in most cases. A shift occurs due to the selfenergy term. 
In the critical region, the eigenvalue
trajectories are exchanged usually. Exceptions occur when 
the exchange is prevented due to certain symmetries, such that one of
the states acts as an 'observer'.

The results obtained at large distances from the critical
region are  true from a mathematical point of view. They are, however, 
hardly realized in physical systems for the following reason. In the
critical region, the system is split into  different parts due to
width bifurcation: one part contains the long-lived states while 
another one contains the short-lived states. The short-lived states decay
quickly, and the long-lived states form a new system the wavefunctions of
which are strongly mixed. This process occurring in the critical
region, is irreversible. As a consequence, the system will never reach the
parameter region far from the critical value $a_{\rm cr}$.  Instead, a
dynamical phase transition takes place as discussed in \cite{top,rotime}.
The  figures \ref{fig4} to \ref{fig9} show that signatures of the
dynamical phase transition can be seen already in the case of only four
overlapping resonances.

\section{Conclusions}

In the present paper, we have shown numerical results 
for avoided  crossings of two and four states in a relatively large
range of the control parameter $a$ by using the matrix 
(\ref{int1}).   The avoided crossing of two levels at the critical
value $a_{\rm cr}$
influences  the complex eigenvalues of the non-Hermitian operator
$\ch$ in a small parameter range around  $a_{\rm cr}$.
Here, the levels  avoid crossing in energy
while the widths cross freely or they cross freely in energy  
while the widths bifurcate. In the first case, the two states are 
exchanged, including their populations. The results can be understood 
by means of the eigenvalue equations (\ref{int6}).

The influence of more than one avoided level crossing occurring 
in a small critical parameter range, onto the eigenvalues of $\ch$ is
stronger than in the case with only one avoided level crossing
in the critical region. The influence 
can be seen still beyond  the critical region.
The imaginary part of the interaction $\omega$ of the states via the
environment affects  the eigenvalues stronger than the real
part of $\omega$ does. Especially width
bifurcation appearing at nearby  avoided level
crossings is usually strong. The reason for these results
are the nonlinear source terms that
appear around an avoided level crossing in the Schr\"odinger equation
\cite{top,rotime}.
They become more important when several avoided level crossings 
are near to one another and the wavefunctions of the different states
overlap. An exception from these results occurs when one state is
related symmetrically to two neighboring states. In such a case, the
state does (almost) not participate in the avoided level crossing
phenomenon, but is solely an 'observer' of the avoided crossing of the
two neighbored states. This result corresponds to sensivity of
exceptional points to the distortion of their symmetry, 
see the discussion in appendix E in \cite{rotime}.  

In physical systems $\omega$ as well as $\gamma_i$ are usually functions of
energy. Furthermore, the ratio Im$(\omega)$ / Re$(\omega)$ increases
with energy.  
We did not simulate these effects in our numerical calculations since 
we are interested in the study of  generic effects. 
In any case,  the critical range of nearby avoided crossings 
affects the eigenstates of a non-Hermitian operator in a 
relatively large parameter range. Far from the critical region with 
overlapping resonances, the eigenvalue trajectories $\ce_i = E_i -
i/2~\Gamma_i$ approach, as it must be from a mathematical point of view, 
the trajectories $\varepsilon_j = e_j -
i/2~\gamma_j$  with $i\ne j$ as a rule. The selfenergy term
causes some shift of the $\ce_i$ relative to the $\varepsilon_j$.

The results shown in figures \ref{fig4} to \ref{fig9}  are relevant 
for physical systems at high level density where different avoided crossings
overlap. This range is characterized by level repulsion and width
bifurcation as the results of the present paper show. Width
bifurcation limits the existence of the system:  it causes a 
splitting of the system into different parts with significant
different lifetimes.  Finally, a dynamical phase transition takes
place \cite{top,rotime}. Further studies
with more than four overlapping resonance states and with $\omega =
\omega (a)$ are in progress.   
Furthermore, we will apply the results of multi-level avoided
crossing  also to the case of atomic and semiconductor
cavity QED which is studied in \cite{appl1,appl2,appl3}.

\end{document}